\documentstyle[12pt,amssymb]{article}
\begin{document}
\font\frak=eufm10 scaled\magstep1
\font\fak=eufm10 scaled\magstep2
\font\fk=eufm10 scaled\magstep3
\font\scriptfrak=eufm10
\font\tenfrak=eufm10


\newtheorem{theorem}{Theorem}
\newtheorem{corollary}{Corollary}
\newtheorem{proposition}{Proposition}
\newtheorem{definition}{Definition}
\newtheorem{lemma}{Lemma}
\def\Eq#1{{\begin{equation} #1 \end{equation}}}

\def\la#1{\lambda_{#1}}
\def\teet#1#2{\theta [\eta _{#1}] (#2)}
\def\tede#1{\theta [\delta](#1)}
\def\N{{\frak N}}
\def\Wei{\wp}

\font\frak=eufm10 scaled\magstep1

\def\bra#1{\langle#1|}
\def\ket#1{|#1\rangle}
\def\goth #1{\hbox{{\frak #1}}}
\def\<#1>{\langle#1\rangle}
\def\cotg{\mathop{\rm cotg}\nolimits}
\def\wt{\widetilde}
\def\const{\hbox{const}}
\def\grad{\mathop{\rm grad}\nolimits}
\def\Div{\mathop{\rm div}\nolimits}
\def\braket#1#2{\langle#1|#2\rangle}
\def\Erf{\mathop{\rm Erf}\nolimits}
\def\cinfty#1{C^\infty(#1)}
\def\Op{\mathop{\rm Op}\nolimits}
\def\Hil{{\cal H}}
\def\Der{\mathop{\rm Der}\nolimits}


\newtheorem{Theo}{Theorem}
\newtheorem{ex}{Example}
\newenvironment{pf}{{\noindent{\it Proof.-}}}{\ $\Box$\medskip}



\mathchardef\za="710B  
\mathchardef\zb="710C  
\mathchardef\zg="710D  
\mathchardef\zd="710E  
\mathchardef\zve="710F 
\mathchardef\zz="7110  
\mathchardef\zh="7111  
\mathchardef\zvy="7112 
\mathchardef\zi="7113  
\mathchardef\zk="7114  
\mathchardef\zl="7115  
\mathchardef\zm="7116  
\mathchardef\zn="7117  
\mathchardef\zx="7118  
\mathchardef\zp="7119  
\mathchardef\zr="711A  
\mathchardef\zs="711B  
\mathchardef\zt="711C  
\mathchardef\zu="711D  
\mathchardef\zvf="711E 
\mathchardef\zq="711F  
\mathchardef\zc="7120  
\mathchardef\zw="7121  
\mathchardef\ze="7122  
\mathchardef\zy="7123  
\mathchardef\zf="7124  
\mathchardef\zvr="7125 
\mathchardef\zvs="7126 
\mathchardef\zf="7127  
\mathchardef\zG="7000  
\mathchardef\zD="7001  
\mathchardef\zY="7002  
\mathchardef\zL="7003  
\mathchardef\zX="7004  
\mathchardef\zP="7005  
\mathchardef\zS="7006  
\mathchardef\zU="7007  
\mathchardef\zF="7008  
\mathchardef\zW="700A  

\newcommand{\be}{\begin{equation}}
\newcommand{\ee}{\end{equation}}
\newcommand{\ra}{\rightarrow}
\newcommand{\lra}{\longrightarrow}
\newcommand{\bea}{\begin{eqnarray}}
\newcommand{\eea}{\end{eqnarray}}
\newcommand{\beas}{\begin{eqnarray*}}
\newcommand{\eeas}{\end{eqnarray*}}
\newcommand{\Z}{{\Bbb Z}}
\newcommand{\R}{{\Bbb R}}
\newcommand{\C}{{\Bbb C}}
\newcommand{\1}{{\bold 1}}
\newcommand{\SL}{SL(2,\C)}
\newcommand{\Sl}{sl(2,\C)}
\newcommand{\SU}{SU(2)}
\newcommand{\su}{su(2)}
\newcommand{\SB}{SB(2,\C)}
\newcommand{\Sb}{sb(2,\C)}
\newcommand{\G}{{\goth g}}
\newcommand{\D}{{\rm d}}
\newcommand{\de}{\,{\stackrel{\rm def}{=}}\,}
\newcommand{\we}{\wedge}
\newcommand{\nn}{\nonumber}
\newcommand{\ot}{\otimes}
\newcommand{\s}{{\textstyle *}}
\newcommand{\ts}{T^\s}
\newcommand{\da}{\dagger}
\newcommand{\pa}{\partial}
\newcommand{\ti}{\times}
\newcommand{\A}{{\cal A}}
\newcommand{\li}{{\cal L}}
\newcommand{\ka}{{\Bbb K}}
\newcommand{\find}{\mid}

\title{Quantum Bi-Hamiltonian Systems}


\author{Jos\'e F. Cari\~nena\\
Depto. F\'{\i}sica Te\'orica, Univ. de Zaragoza\\
50009 Zaragoza, Spain\\
{\it e-mail:} jfc@posta.unizar.es
\and
Janusz Grabowski\thanks{Supported by KBN, grant No. 2 P03A 031 17.}\\
Institute of Mathematics, Warsaw University\\
ul. Banacha 2, 02-097 Warszawa, Poland. \\
and\\
Mathematical Institute, Polish Academy of Sciences\\
ul. \'Sniadeckich 8, P. O. Box 137, 00-905 Warszawa, Poland\\
{\it e-mail:} jagrab@mimuw.edu.pl
\and
Giuseppe Marmo\\
Dipartimento di Scienze Fisiche,
Universit\`a Federico II di Napoli\\
and\\
INFN, Sezione di Napoli\\
Complesso Universitario di Monte Sant'Angelo\\
Via Cintia, 80125 Napoli, Italy\\
{\it e-mail:} marmo@na.infn.it}
\maketitle
\begin{abstract} We define quantum bi-Hamiltonian systems, by analogy  with
the classical case, as derivations in operator algebras which are  inner
derivations with respect to two compatible  associative  structures.  We
find such structures by means of the  associative  version  of  Nijenhuis
tensors. Explicit examples, e.g. for the harmonic oscillator, are given.
\end{abstract}

\section{Introduction}
Bi-Hamiltonian systems at  the  classical  level,  as  noticed  by
F.~Magri (\cite{Ma}), play an important role  in   the   discussion   of
complete integrability in the sense of Liouville.
\par
At the  quantum  level,  much  earlier,  E.~P.~Wigner  raised  the
question: {\it Do the equations of motion  determine  the  quantum
mechanical commutation relations?}
\par
The way Wigner formulated his question was the following. Assuming
the equations of motion
\be
{i}\frac{\D}{\D t}\hat q=\frac{\hat p}{m},\qquad
{i}\frac{\D}{\D t}\hat p=-\frac{\widehat{\pa V}}{\pa q},
\ee
to find commutation relations such that
\be
\frac{\D}{\D t}\hat q=-\frac{i}{\hbar}[\hat q,\hat H],\qquad
\frac{\D}{\D t}\hat p=-\frac{i}{\hbar}[\hat p,\hat H].
\ee
Wigner argued that equations  of  motion  have  a  more  immediate
physical significance than the canonical commutation relations
\be
[\hat p,\hat q]=-i\hbar.
\ee
The commutation relations we are searching  for  should  define  a
`quantum Poisson bracket' in  the   terminology   of   Dirac \cite{Di}.
Indeed, Dirac shows that if we look for a Lie  algebra  structure  on
the space of observables such that
\be\label{1}
[A,BC]=[A,B]\,C+B\,[A,C],
\ee
then necessarily
\be
[A,B]=\zl\, (AB-BA),
\ee
with $\zl$ being any complex number.
To put it differently, according to Dirac, to look for  alternative
commutation relations (with the additional requirement (\ref{1})),
it is equivalent to look for alternative products on the space  of
observables with the requirement  that  the  equations  of  motion
define a derivation with respect to the associative product.
\par
Recently it has been  shown (\cite{MMSZ}),  in  connection   also   with
deformed oscillators, that one may obtain  a  large  class  of
alternative associative products of the kind
\be\label{0}
A\circ_KB=AKB
\ee
for which the dynamics is a derivation any time $K$ is an observable which is 
a  constant
of the motion. In particular,  it  has  been  applied  to  a  precessing
magnetic dipole \cite{LPMM}.
It  turns  out  that  all  these  deformations  are
compatible  among  themselves in the sense we will explain later.  This  is  
rather   unsatisfactory,
because in considering the classical limit of these quantum  cases
we should be able to recover  Poisson  structures  which  are  not
necessarily compatible.
\par
This  note  is  an  attempt  to  put  the  search  of  alternative
associative products in a more systematic setting.

\section{Some important concepts in cohomology of algebras}

  Let $({\cal A},*)$  be an associative algebra and  $V$ be a 
   ${\cal A}$--bimodule, respectively. In other
 words, $V$ is a module that is the carrier space for a linear
 representation $\Psi$ of  ${\cal A}$ and a linear antirepresentation $\Psi'$
 of ${\cal A}$ that commute.
  
 By a   $n$--cochain  we mean a $n$--linear  
 mapping from ${\cal A}\times\dots\times{\cal A}$ ($n$~times) into $V$.  
 We denote by 
 $C^n({\cal A},V)$ 
 the space of such $n$--cochains that can be regarded as an additive group. 
 For every $n\in\N$ we introduce the Hochschild (\cite{GH}) coboundary
operator, as defined by 
Eilenberg and  Mac Lane,
 $\delta:C^n({\cal A},V)\to
 C^{n+1}({\cal A},V)$, by  
 \begin{eqnarray}\nn
 (\delta\alpha)(a_1,\dots,a_{n+1}) 
  &=& a_1\alpha(a_2,\dots,a_{n+1})+ \sum_{i=1}^{n} (-1)^{i} 
        \alpha(a_1,\dots,   a_i* a_{i+1},\dots,a_{n+1})+  \\
  &+&  (-1)^{n+1}\alpha( a_1,\dots,\dots,a_n)a_{n+1}\ . 
 \end{eqnarray}

  It is now easy to check that 
 $$
 \delta\circ \delta=0\ .
 $$

 The cohomology groups can be defined as follows: an $n$ cochain
$\alpha\in  C^n({\cal A},V)$ is called an $n$--cocycle
 if $\delta\alpha=0$, and an element of the form $\delta\beta$
 where $\beta \in C^{n-1}({\cal A},V)$ is called an $n$--coboundary. These 
form 
 a subgroup $B^n({\cal A},V)$ of the additive group $Z^n({\cal A},V)$ of 
 $n$--cocycles. The cohomology group $H^n({\cal A},V)$    is defined as the 
 quotient group $H^n({\cal A},V)=Z^n({\cal A},V)/B^n({\cal A},V)$.
 
 For instance,  when $n=1$, we obtain
 $$(\delta \alpha_1)(a_1,a_2)= a_1 \alpha_1(a_2) -\alpha_1(a_1*a_2) 
 +\alpha_1(a_1)a_2\ , 
 $$
 and for $n=2$,
 $$
 (\delta \alpha_2)(a_1,a_2,a_3)=a_1 \alpha_2(a_2,a_3) -\alpha_2(a_1*a_2,a_3) 
 +\alpha_2(a_1,a_2*a_3) -\alpha_2(a_1,a_2) a_3\,. 
 $$
 
 The simplest example obtains when $V$ is the additive group of $\cal A$,
and then 
 the $\cal A$--bimodule structure is given by left and right
multiplication.

\section{Compatible associative products and associative  Nijenhuis
tensors}
By analogy with the classical case, where a bi-Hamiltonian system
consists of two compatible Poisson brackets and a system which  is
Hamiltonian with respect to both brackets, by a {\em weak quantum
bi-Hamiltonian  system}  we  shall mean two Lie algebra
structures on the space $\Op(\Hil)$ of operators on a Hilbert
space $\Hil$ (one of them will be usually the original one) which are
compatible in the sense that the corresponding commutators are
compatible Lie brackets (i.e. their sum is again a Lie bracket) and
a derivation $D\in \Der(\Op(\Hil))$ which is an inner derivation with
respect to both associative structures \cite{DMS}.
\par
Since we want the Leibniz rule 
\be
[A,B\circ C]=[A,B]\circ C+B\circ[A,C],
\ee
in view of the Dirac's proof  (\cite{Di}, pp. 85-86),  that  derivations
of a
sufficiently  non-degenerate  associative  algebra  are  just  adjoint
operators, we would like to  have  a  new  bracket  in   the   form   of
the commutator of a  new associative structure. We will call such
pairs of associative structures just {\em weak quantum
bi-Hamiltonian} ones. A  possible additional requirement  is  that   both
associative structures have the same unit $\1$. Let us note that one can
also consider a stronger version of compatibility of associative
products ``$\circ_1$" and ``$\circ_2$" requiring that $\circ_1+\zl\circ_2$
is associative for all $\zl\in\ka$, where  $\Bbb K$  is  the  ground  field
(then the mean $(\circ_1+\circ_2)/2$ is again associative with the same
unit $\1$) and this is what we mean  by  a  {\em  quantum  bi-Hamiltonian
system}. We start with some pure algebraic observations.
\par
Let $(\A,\,\cdot\,)$ be a unital associative algebra. A simple way  to
define a new associative product on $\A$ is to take an element
$K\in\A$ and to define a new product by
\be
A\circ_KB=AKB.
\ee
(We will usually skip the product symbol for the original associative
structure.)
Observe that the unit is not preserved unless $K=\1$ and that   we
have the homomorphism of the products
\be
T_K(A\circ_KB)=T_K(A)T_K(B)
\ee
for $T_K$ being the linear map
\be
T_K:\A\ra\A,\quad T_K(A)=KA,
\ee
which is an isomorphism (non-unital, however) in case
$K$ is invertible.
\par
This example can be  generalized  if  we  deform  the  associative
structure by an associative analog of the Nijenhuis  map  (tensor),
known better in the Lie algebra case.

\medskip\noindent

Let $(\A,\mu)$ be an associative algebra over  a  field
$\ka$, with the product
\be
\zm:\A\times\A\ra\A,\qquad (A, B)\mapsto AB
\ee
and let $N:\A\ra\A$ be a linear map ($N\in \A^*\otimes\A$). If $N$ is a 
derivation 
of the algebra $(\A,\mu)$, then
$N(A)B+AN(B)-N(AB)=0$. In any case, the map 
\be\label{2}
\mu_N:(A,B)\mapsto A\circ_NB=N(A)B+AN(B)-N(AB)\ ,
\ee
is a bilinear map and therefore it defines a new algebra structure  
$(\A,\mu_N)$.
Using the terminology introduced in the preceding section, and considering 
the $\cal A$--bimodule structure in  $\cal A$ as given by left and right 
multiplication, we can say that $A\circ_NB=\delta_\mu N(A,B)$ and
therefore that 
$N$ is a derivation of the original algebra if and only if $N$ is a
1-cocycle with respect to the Hochschild coboundary operator $\zd_\zm$
associated with the product $\zm$.

The obstruction for the linear map $N$ to be a homomorphism of these
products is measured by the $\zm$-Nijenhuis torsion of $N$:
\be
T_N(A,B)=N(A\circ_NB)-N(A)N(B)\ .
\ee    

\begin{definition}
We say
that the linear map $N:\A\to\A$ is a  $\zm$-{\it Nijenhuis  tensor} if the
 $\zm$-Nijenhuis torsion of $N$ vanishes, $T_N(A,B)=0, \ \forall A,B\in \A$.
\end{definition}
 In this case $N$ is a homomorphism of the corresponding products:
\be
N(A\circ_NB)=N(A)N(B)\ ,
\ee
i.e.
$$N(N(A)B+AN(B)-N(AB))-N(A)N(B)=0\ .
$$

\begin{theorem} The product $\zm_N$ defined by (\ref{2}) is associative if
and only if the $\zm$-Nijenhuis torsion $T_N$ of $N$ is a 2-Hochschild
cocycle of the algebra $\A$, i.e.
\be
\zd_\zm T_N(A,B,C):=AT_N(B,C)-T_N(AB,C)+T_N(A,BC)-T_N(A,B)C=0.
\ee
If this is the case, $\zm_N$ is an associative product compatible with
$\zm$, i.e.  $\zm+\zl\zm_N$  are  associative for all $\zl\in{\ka}$.
If $\zm$ is unital with the unit  $\1$, then $\zm_N$ has the same unit
providing that  $N(\1)=\1$.
\par
In particular, if
$N$  is  a   $\zm$-Nijenhuis   tensor,   then $\zm_N$ is an
associative product on $\A$ which is compatible with $\zm$.
\end{theorem}
\begin{pf} By direct computation,
\beas
&(A\circ_NB)\circ_NC-A\circ_N(B\circ_NC)=\\
&-AT_N(B,C)+T_N(AB,C)-T_N(A,BC)+T_N(A,B)C=-\zd_\zm T_N(A,B,C).
\eeas
As   for   the compatibility, it suffices to prove
\be\label{comp}
(AB)\circ_NC+(A\circ_NB)C=A(B\circ_NC)+A\circ_N(BC),
\ee
which is straightforward:
\beas
(AB)\circ_NC+(A\circ_NB)C&=&N(AB)C+ABN(C)-N(ABC)\\
&&+N(A)BC+AN(B)C-N(AB)C\\
&=&N(A)BC+AN(B)C+ABN(C)-N(ABC)\\
&=&A(B\circ_NC)+A\circ_N(BC).
\eeas
\hskip15cm\end{pf}

\medskip\noindent
The relation  (\ref{comp}) means that the map $\mu_N$, as seen as
2-cochain in the algebra $(\A,\mu)$, is a 2-cocycle because 
$$\delta_\mu{\mu_N}(A,B,C)=A(B\circ_NC)-(AB)\circ_NC+A\circ_N(BC)
-(A\circ_NB)C.$$

\medskip\noindent
{\bf Remark.} Note that the compatibility condition (\ref{comp}) holds
automatically, no matter if $\zm_N$ is associative or not.  If  we  look
for a new associative product $\circ$ which is compatible in  the  sense
of (\ref{comp}), then this means that the new product  is  a  Hochschild
cocycle of the original associative algebra.  If  our  algebra  is,  for
instance, the algebra  of  $n\times  n$  matrices,  due  to  the  Morita
equivalence (cf. \cite{Lo}), its Hochschild cohomology are the
same as the Hochschild cohomology of $\ka$ (regarded as
1-dimensional algebra over itself), thus vanish in dimensions higher than
zero, so our
product $\circ$ has to be a Hochschild coboundary, i.e. of the form
$\circ_N$  for  some  $N$.
This shows that we have not much freedom and, looking for compatible
associative products, we must, in principle, work with Nijenhuis tensors.

\medskip
The above observations can be reformulated in terms of the
so called {\it Gerstenhaber bracket} $[\ ,\ ]_G$, which is a
graded Lie bracket on the graded space of multilinear maps of $\A$ into
$\A$ and which recognizes associative products (cf. \cite{Ge, Gr}), in
full correspondence with the analogous theory
for Nijenhuis tensors for Lie algebras and the Richardson-Nijenhuis bracket
(cf. \cite{KSM}). In particular,
$\zm_N=[\zm,N]_G$ and
\be
2T_N(A,B)=[N,[\zm,N]_G]_G+[\zm,N^2]_G,
\ee
so that the Theorem 1 is a direct consequence of the graded Jacobi
identity for the Gerstenhaber bracket and the fact that $[\zm,\cdot]_G$ is
proportional to the Hochschild coboundary operator $\zd_\zm$ (In
particular, $[\zm,\zm]_G=0$, so that $[\zm,[\zm,N]_G]_G=0$, which is the
compatibility condition (\ref{comp}).)

\medskip\noindent
Now, we will show that a Nijenhuis tensor gives rise to a whole hierarchy 
of Nijenhuis tensors and associative structures, as has been already
discovered by Saletan \cite{Sa}. Putting $N^k$ instead of $N$ in the 
above, we can consider products $\zm_{N^k}$.
\begin{lemma} If $N$ is a $\zm$-Nijenhuis tensor, then the products
$\zm_{N^{k+r}}$ and $\zm_{N^k}$ are $N^r$-related, i.e.
\be\label{s1}
N^r(A\circ_{N^{k+r}}B)=N^r(A)\circ_{N^k}N^r(B)
\ee
for all $k,r=0,1,2,\dots$.
\end{lemma}
\begin{pf} We will start with proving
\be\label{s2}
N(A\circ_{N^{k+1}}B)=N(A)\circ_{N^k}N(B).
\ee
Applying $N^k$ to the Nijenhuis torsion
\be
N(N(A)B)+N(AN(B))-N^2(AB)-N(A)N(B),
\ee
which vanish by assumption, we get
\be\label{s3}
N^{k+2}(AB)-N^{k+1}(AN(B))=N^{k+1}(N(A)B)-N^k(N(A)N(B)).
\ee
Using (\ref{s3}) inductively for $k:=k-1$, we end up with
\be\label{s4}
N^{k+2}(AB)-N^{k+1}(AN(B))=N(N^{k+1}(A)B)-N^{k+1}(A)N(B).
\ee
In a similar way, we get
\be
N^{k+2}(AB)-N^{k+1}(N(A)B)=N(AN^{k+1}(B))-N(A)N^{k+1}(B)
\ee
which, combined with (\ref{s3}), gives
\be\label{s5}
N^{k+1}(AN(B))-N^k(N(A)N(B))=N(AN^{k+1}(B))-N(A)N^{k+1}(B).
\ee
Combining now (\ref{s5}) and (\ref{s4}), we find
\beas
&&N^{k+2}(AB)-N^k(N(A)N(B))=\\
&&N(N^{k+1}(A)B+AN^{k+1}(B))-N^{k+1}(A)N(B)
-N(A)N^{k+1}(B)
\eeas
which can be rewritten in the form
\beas
&&N^k(N(A))N(B)+N(A)N^k(N(B))-N^k(N(A)N(B))=\\
&&N(N^{k+1}(A)B+AN^{k+1}(B)-N^{k+1}(AB)).
\eeas
But the last one is exactly (\ref{s2}).
Now, applying (\ref{s2}) inductively
\be
N^r(A\circ_{N^{k+r}}B)=N^{r-1}N(A\circ_{N^{k+r}}B)=
N^{r-1}(N(A)\circ_{N^{k+r-1}}N(B)),
\ee
we end up with (\ref{s1}).

\smallskip\noindent\hskip15cm
\end{pf}
\begin{theorem} If $N$ is a $\zm$-Nijenhuis tensor, then
\be\label{ss1}
(\zm_{N^i})_{N^k}=\zm_{N^{i+k}}
\ee
and $N^r$ is a $\zm_{N^i}$-Nijenhuis tensor, i.e.
\be
N^r(A(\circ_{N^i})_{N^r}B)=N^r(A)\circ_{N^i}N^r(B)
\ee
for    all    $i,k=0,1,2,\dots$.  In    particular,    
all products $\zm_{N^k}$ are associative and compatible.
\end{theorem}
\begin{pf} First, 
we show that 
\be\label{ss2}
(\zm_{N^i})_N=\zm_{N^{i+1}}.
\ee
Indeed, 
\beas
A(\circ_{N^i})_NB&=&N(A)\circ_{N^i}B+A\circ_{N^i}N(B)-N(A\circ_{N^i}B)\\
&=&N^{i+1}(A)B+N(A)N^i(B)-N^i(N(A)B)+N^i(A)N(B)\\
&&+AN^{i+1}(B)-N^i(AN(B))
-N(A)\circ_{N^{i-1}}N(B)\\
&=&N^{i+1}(A)B+AN^{i+1}(B)-N^{i+1}(AB)-N^i(N(A)B\\
&&+AN(B)-N(AB))+N^{i-1}
(N(A)N(B))\\
&=&N^{i+1}(A)B+AN^{+1}(B)-N^{i+1}(AB)\\
&&-N^{i-1}(N(A\circ_NB)-N(A)N(B))
=A\circ_{N^{i+1}}B,
\eeas
where we have used, according to Lemma,
$N(A\circ_{N^i}B)=N(A)\circ_{N^{i-1}}N(B)$.
Now, (\ref{ss2}) together with (\ref{s2}) show that $N$ is a 
$\zm_{N^i}$-Nijenhuis tensor which produces a compatible associative
product $(\zm_{N^i})_N=\zm_{N^{i+1}}$.
Thus we can apply Lemma and (\ref{ss2}) to $\zm_{N^i}$ instead of  $\zm$
that proves the theorem.

\smallskip\noindent\hskip15cm
\end{pf}

\medskip\noindent
There is a way to obtain a new Nijenhuis tensor from two of them.
Two Nijenhuis tensors $N_1$ and $N_2$ on $\A$ will be said  to  be  {\it
compatible} if $N_1+N_2$ is again a Nijenhuis tensor.
\begin{theorem}
Nijenhuis tensors $N_1$ and $N_2$ are compatible if and only if
\be\label{com}
N_1(A\circ_{N_2}B)+N_2(A\circ_{N_1}B)=N_1(A)N_2(B)+N_2(A)N_1(B).
\ee
If $N_1$ is compatible with $N_2,\dots,N_k$, then it is compatible
with any linear combination of them. If $N_1,\dots,N_k$ are pairwise
compatible, then any two linear combinations of them are compatible.
\end{theorem}
\begin{pf} The firs statement is a direct consequence of definitions 
if we only observe that $\circ_{N_1+N_2}=\circ_{N_1}+\circ_{N_2}$. 
The rest follows from the fact that (\ref{com}) depends linearly on
$N_1$ and $N_2$.

\smallskip\noindent\hskip15cm
\end{pf}
\begin{theorem}
If $N$ is a  Nijenhuis  tensor, then all linear 
combinations of $N^k$, $k=0,1,2,\dots,$  are compatible.
\end{theorem}
\begin{pf}
Indeed, for $k\ge r$,
\beas
&&N^k(A\circ_{N^r}B)+N^r(A\circ_{N^k}B)=\\
&&N^{k-r}(N^r(A)N^r(B))+N^r(A)\circ_{N^{k-r}}N^r(B)=\\
&&N^k(A)N^r(B)+N^r(A)N^k(B),
\eeas
where we have used Theorem 2. Now, the assertion follows from Theorem 3.

\smallskip\noindent\hskip15cm
\end{pf}

\medskip\noindent
{\bf Remark.} Let us observe that the product (\ref{0}) can be  obtained
from the Nijenhuis tensor $N_K(A)=KA$. Indeed,
\be
A\circ_{N_K}B=(KA)B+A(KB)-K(AB)=AKB.
\ee
The operator $N_K$ is a Nijenhuis tensor, since
\be
N_K(A\circ_{N_K}B)=K(AKB)=(KA)(KB)=N_K(A)N_K(B).
\ee
In particular, the operators of multiplication by elements of the  field
$\ka$ are Nijenhuis tensors. Other examples of Nijenhuis tensors can
be constructed in the following way.

\begin{theorem}  If  $\A=\A_1\oplus\A_2$  is  a  decomposition   of   an
associative algebra $\A$ (nonunital in general) with the multiplication
$\zm$ into  two subalgebras ($\A$ with such a decomposition is called
{\it a twilled algebra})
and $P_1,P_2$ denote the corresponding projections of $\A$  onto  $\A_1$
and  $\A_2$,  respectively,  then  any  linear  combination  $N=\zl_1P_1
+\zl_2P_2$ is a $\zm$-Nijenhuis tensor.
\end{theorem}
\begin{pf}  Since  $\zl_1P_1+\zl_2P_2=(\zl_1-\zl_2)P_1+\zl_2I$,  it   is
sufficient to show that $P_1$ is a  $\zm$-Nijenhuis  tensor.  Using  the
decomposition $A=A_1+A_2$ etc., we have
\be
A\circ_{P_1}B=A_1B+AB_1-(AB)_1,
\ee
so that $\zm_{P_1}=\zm$ on $\A_1$, $\zm_{P_1}=0$ on $\A_2$, and
\be
A_1\circ_{P_1}B_2=P_2(A_1B_2),\qquad A_2\circ_{P_1}B_1=P_2(A_2B_1).
\ee
Hence,
\be\label{3}
A\circ_{P_1}B=A_1B_1+P_2(A_1B_2+A_2B_1)
\ee
and
\be
P_1(A\circ_{P_1}B)=P_1(A_1B_1+P_2(A_1B_2+A_2B_1))=A_1B_1=P_1(A)P_1(B),
\ee
so that $P_1$ is a Nijenhuis tensor.

\smallskip\noindent\hskip15cm
\end{pf}

\medskip\noindent
{\bf Example 1.} Take $\A=M_2({\ka})$  --  the  algebra  of  $2\ti
2$-matrices  $A=\left(\begin{array}{cc}a&b\\c&d\end{array}\right)$.
Take $\A_1$ to be the algebra of upper-triangular matrices
$A=\left(\begin{array}{cc}a&b\\0&d\end{array}\right)$ and let $\A_2$  be
the (commutative) algebra of strictly lower-triangular matrices
$A=\left(\begin{array}{cc}0&0\\c&0\end{array}\right)$.    Taking     the
Nijenhuis tensor $P_1$, we get new associative matrix multiplication  in
the form
\be
\left(\begin{array}{cc}a&b\\c&d\end{array}\right)\circ
\left(\begin{array}{cc}a'&b'\\c'&d'\end{array}\right)=
\left(\begin{array}{cc}aa'&ab'+bd'\\ca'+dc'&dd'\end{array}\right).
\ee
Note that the unit matrix $I$ remains the unit for this new product  and
that inner derivations given by diagonal matrices are the same for  both
products.
\par
Of course, we can use the complementary projection instead and  get  the
product
\be
\left(\begin{array}{cc}a&b\\c&d\end{array}\right)\circ'
\left(\begin{array}{cc}a'&b'\\c'&d'\end{array}\right)=
\left(\begin{array}{cc}bc'&0\\0&cb'\end{array}\right)
\ee
which is associative but not unital. Of course, this example admits an
obvious generalization to algebras of matrices of any dimension.
Note also that  we  can  consider  these
products at the level of the operator algebra  $\Op(\Hil)$  over  a
Hilbert
space $\Hil$ directly, viewing this algebra as algebra of infinite matrices,
or using a decomposition of $\Hil$ into a direct sum of
subspaces, so that we can write operators in a matrix form.

\bigskip\noindent
{\bf Remark.} Observe that the product (\ref{3})
is  associative  even  if  $\A_2$  is  not  a  subalgebra  but  just   a
complementary subspace. Indeed,
\beas
(A\circ B)\circ C&=&(A_1B_1+P_2(A_1B_2+A_2B_1))\circ C\\
&=&A_1B_1C_1+P_2(A_1B_1C_2+P_2(A_1B_2+A_2B_1)C_1)\\
&=&A_1B_1C_1+P_2(A_1B_1C_2+A_1B_2C_1+A_2B_1C_1)=A\circ (B\circ C).
\eeas
However, this product is not of the  form  $\zm_{P_1}$  and  it  is,  in
general, not compatible with the original one.

\medskip\noindent
{\bf Example 2.} Again, for the matrix algebra $\A=M_2({\ka})$  take
$\A_1=span<I,C>$, $\A_2=span<A,B>$, where
\be\nn
A=\left(\begin{array}{cc}1&0\\0&-1\end{array}\right),\quad
B=\left(\begin{array}{cc}0&1\\1&0\end{array}\right),\quad
C=\left(\begin{array}{cc}0&1\\-1&0\end{array}\right),\quad
I=\left(\begin{array}{cc}1&0\\0&1\end{array}\right).
\ee
Using the Nijenhuis tensor $P_1$, we get the product
\beas
A\circ B&=&B\circ A=0,\qquad A\circ A=0,\qquad B\circ B=0,\\
A\circ C&=&B,\qquad C\circ A=-B,\\
B\circ C&=&-A,\qquad C\circ B=A,\qquad C\circ C=-I,
\eeas
and $I$  remains  the  unit  for  this  product.  The  inner  derivation
associated with $C$ is the same for both products.

\bigskip
The product (\ref{3}) is in fact a contraction of the
original one, since
\be
A\circ_{P_1}B=\lim_{h\ra 0}T_h^{-1}(T_h(A)T_h(B)),
\ee
where $T_h(A)=A_1+hA_2$. Indeed,
\beas
T_h^{-1}(T_h(A)T_h(B))&=&T_h^{-1}(A_1B_1+h(A_2B_1+A_1B_2)+h^2A_2B_2)\\
&=&A_1B_1+P_2(A_2B_1+A_1B_2)+hP_1(A_2B_1+A_1B_2)\\
&&+hP_2(A_2B_2)+h^2P_1(A_2B_2)
\eeas
which tends to $A_1B_1+P_2(A_2B_1+A_1B_2)$ as $h\ra 0$.
\par
This can be generalized as follows. Using a decomposition of the algebra
$\A$ into the direct sum $\A=\A_1\oplus\A_2$, where $\A_1$ is assumed to be
a subalgebra we will write $A=A_1+A_2$ for any element $A\in\A$ accordingly
to this decomposition. Suppose that we have invertible linear operators
$N_1,N_2$ acting, respectively, on $\A_1$ and $\A_2$. For any $h\in\ka$ we
define $T_h:\A\ra\A$ by $T_h(A)=N_1(A_1)+hN_2(A_2)$ and put
\be
A\circ_hB=T_h^{-1}(T_h(A)T_h(B)).
\ee
The product ``$\circ_h$'' is clearly associative and
\beas
A\circ_hB&=&N_1^{-1}(N_1(A_1)N_1(B_1))+\\
&&N_2^{-1}((N_1(A_1)N_2(B_2)+N_2(A_2)N_1(B_1)+hN_2(A_2)N_2(B_2))_2)+\\
&&hN_1^{-1}((N_1(A_1)N_2(B_2)+N_2(A_2)N_1(B_1)+hN_2(A_2)N_2(B_2))_1).
\eeas
Passing formally with $h\ra0$, we get the contracted associative product
\be\label{prod1}
A\circ B=N_1^{-1}(N_1(A_1)N_1(B_1))+
N_2^{-1}((N_1(A_1)N_2(B_2)+N_2(A_2)N_1(B_1))_2).
\ee
If we assume that there is an associative product $\circ_1$ on $\A_1$
such that $N_1(A_1\circ_1B_1)=N_1(A_1)N_1(B_1)$, then we can write
the product (\ref{prod1}) in the form
\be\label{prod2}
A\circ B=N_1(A_1)\circ_1N_1(B_1)+
N_2^{-1}((N_1(A_1)N_2(B_2)+N_2(A_2)N_1(B_1))_2).
\ee
Now we can skip the assumption that $N_1$ is invertible. We can get even
more, as one can check by direct calculations.
\begin{theorem} Let $\A=\A_1\oplus\A_2$ be a decomposition of an
associative algebra into subspaces such that $\A_1$ is a subalgebra. Let
$\circ_1$ be an additional associative product on $\A_1$ and let
$N_1,N'_1:\A_1\ra\A_1$ be homomorphisms of the product $\circ_1$ into the
original one ($N_1(A_1\circ_1B_1)=N_1(A_1)N_1(B_1)$, etc.). Then, for any
invertible linear map $N_2:\A_2\ra\A_2$, the product
\be\label{prod}
A\circ B=A_1\circ_1B_1+N_2^{-1}((N_1(A_1)N_2(B_2)+N_2(A_2)N'_1(B_1))_2)
\ee
is an associative product on $\A$.
\end{theorem}
We obtain a particular case of the above theorem if we start with a Nijenhuis
tensor $N_1$ on the subalgebra $\A_1$ and we put $N'_1=N_1$ and
$\circ_1=\circ_{N_1}$. 

\medskip\noindent
{\bf Example 3.} Let $\A$ be a matrix algebra, $\A_1$ be the subalgebra
of diagonal matrices, and $\A_2$ be the complementary subspace of
matrix with 0 on the diagonal. Denote by $\zD(A)$ the diagonal part
of the matrix $A$. We define $N_1:\A_1\ra\A_1$ to be the multiplication
by an invertible diagonal matrix $K$ which is a Nijenhuis tensor. 
We have $A\circ_1B=KAB$ for diagonal matrices $A$ and $B$. Finally,
putting $N'_1=N_1$ and $N_2=I$ on $\A_2$, we get a new associative 
product (\ref{prod})
\bea
A\circ B&=&K\zD(A)\zD(B)+K\zD(A)(B-\zD(B))+(A-\zD(A))K\zD(B)=\\ 
\nn &&K\zD(A)B+AK\zD(B)-K\zD(A)\zD(B).
\eea
W have used the fact that $\A_2$ is invariant with respect to the
multiplication by diagonal matrices. Note also, that the above product
is not $\zm_{K\zD}$ since, in general, $\zD(A)\zD(B)\ne\zD(AB)$.

\medskip\noindent
To construct a Nijenhuis tensor $N$ on $\A=\A_1\oplus
\A_2$ from $N_1$, we can use the following.
\begin{theorem}
If $N_1$ is a Nijenhuis tensor on the subalgebra $\A_1$ in the decomposition
$\A=\A_1\oplus\A_2$, then $N(A)=N_1(A_1)$ is a Nijenhuis tensor on $\A$
if and only if
\bea\label{u}
&N_1^2((A_2B_2)_1)=0,\\
&N_1((N_1(A_1)B_2)_1-N_1((A_1B_2)_1))=0,\nn\\
&N_1((A_2N_1(B_1))_1-N_1((A_2B_1)_1))=0,\nn
\eea
far all $A_i,B_i\in\A_i$, $i=1,2.$ In particular, this is the case
for $\A_2$ being a (two-sided) ideal.
\end{theorem}
\begin{pf} Since
\beas
A\circ_NB&=&N_1(A_1)B+AN_1(B_1)-N_1((AB)_1)\\
&=&A_1\circ_{N_1}B_1+N_1(A_1)B_2+A_2N_1(B_1)-
N_1((A_1B_2+A_2B_1+A_2B_2)_1),
\eeas
we just rewrite the condition $N_1((A\circ_NB)_1)=N_1(A_1)N_1(B_1)$
using the fact that $N_1$ is a Nijenhuis tensor and that $A_i,B_i\in\A_i$
can be chosen independently.

\smallskip\noindent\hskip15cm
\end{pf}

\medskip\noindent
{\bf Example 4.} Let $\A$ be the algebra of $n\times n$-matrices which are
upper-triangular. Take $\A_1$ to be the commutative subalgebra of diagonal
matrices and $\A_2$ to be the complementary subalgebra of strictly
upper-triangular matrices. Put $N_1$ to be the multiplication by a diagonal
matrix $K$ from the left. Then $N_1$ is a Nijenhuis tensor on $\A_1$ which
can be extended to the Nijenhuis tensor $N(A)=N_1(A_1)$ on $\A$.
Indeed, in this case $\A_2$ is an ideal. The corresponding deformed
product has the form
\be
A\circ_NB=K\Delta(A)B+AK\Delta(B)-K\Delta(AB),
\ee
where $\Delta(A)$ denotes the diagonal part of $A$.

\bigskip\noindent
Let us recall (cf. \cite{KSM}) that a Nijenhuis tensor for  a  Lie  algebra
$(\li,[\cdot,\cdot])$   is   a   linear   mapping   $N:\li\ra\li$   such
that $N([A,B]_N)=[N(A),N(B)]$, where
\be
[A,B]_N=[N(A),B]+[A,N(B)]-N[A,B].
\ee
It is well known (see e.g. \cite{KSM}) that $[\cdot,\cdot]_N$ is a
compatible Lie
bracket if  $N$  is  a  Nijenhuis  tensor.  The  relation  between
Nijenhuis tensors in the associative and Lie algebra cases describes
the following.
\begin{theorem}
If  $N$  is  a  $\zm$-Nijenhuis  tensor  for  an   associative   algebra
$(\A,\zm)$, then $N$ is a Nijenhuis tensor for the Lie  algebra  $(\A,[\
,\ ])$, where $[A,B]=AB-BA$ is the commutator, and
\be
[A,B]_N=A\circ_NB-B\circ_NA,
\ee
i.e. the deformed Lie bracket $[\cdot,\cdot]_N$  is  the  commutator  of
the deformed associative product $\circ_N$.
\end{theorem}
\begin{pf} By definition,
\beas
[A,B]_N&=&[N(A),B]+[A,N(B)]-N[A,B]\\
&=&N(A)B-BN(A)+AN(B)-N(B)A-N(AB-BA)\\
&=&(N(A)B+AN(B)-N(AB))-(N(B)A+BN(A)-N(BA))\\
&=&A\circ_NB-B\circ_NA.
\eeas
Then,
\be\nn
N([A,B]_N)=N(A\circ_NB-B\circ_NA)=N(A)N(B)-N(B)N(A)=[N(A),N(B)],
\ee
i.e. $N$ is a Lie-Nijenhuis tensor.

\smallskip\noindent\hskip15cm
\end{pf}

The above shows that we can apply the well-known theory of
Nijenhuis tensors in the Lie algebra case for the `commutator
part' of the associative Nijenhuis tensor to construct commuting
elements etc. On the other hand, it is harder to find associative
Nijenhuis tensors, since vanishing of the Nijenhuis torsion in the Lie
algebra case
\be
N(A\circ_NB-B\circ_NA)=N(A)N(B)-N(B)N(A)
\ee
refers only to the skew-symmetrizations (commutators) of the corresponding
products. Similarly, $[A,B]_N=A\circ_NB-B\circ_NA$ is a Lie bracket
if and only if the total skew-symmetrization of the associator
\be
Ass_N(A,B,C)=(A\circ_NB)\circ_NC-A\circ_N(B\circ_NC)
\ee
vanishes, which is weaker than just vanishing of the associator itself.
\section{Final Examples}
{\bf Example 5.} 
Let  now  the  algebra  $\A$  be  the  algebra  of   infinite   matrices
concentrated about the diagonal, i.e. matrices which are  null  outside
a diagonal strip.  The  algebra  $\A$  represents  then unbounded
operators on a Hilbert space $\cal H$ with a  common  dense  domain.  We
choose $\A_1$ to be a subalgebra of upper-triangular  matrices  and  for
$\A_2$ we take the supplementary  algebra  of  strict  lower-triangular
matrices.  Then,  the   mapping
\be
N_\lambda(A)=(1-\lambda)A_1+\lambda A
\ee
is a Nijenhuis tensor  on  $\A$,  in  view of   of Theorem 4,
for  every  $\lambda\in{\Bbb  C}$.  Since  the  corresponding   deformed
associative products $\circ_\lambda$ give all the same result if one  of
factors is a diagonal matrix,   the Hamiltonian $H$ for the harmonic
oscillator, $H(|e_n\rangle)=n| e_n\rangle$, describes the same 
motion for all deformed brackets. This time, however, $a^\da\circ_\lambda  
a=\lambda H$.

\medskip\noindent
{\bf Example 6.} Let us end up with a version of Example 3 for the
algebra $\A$ of unbounded operators as above.
Recall that our deformed product is
\be
A\circ B=K\zD(A)B+AK\zD(B)-K\zD(A)\zD(B),
\ee
where $\zD(A)$ is the diagonal part of $A$. If $A$ is diagonal, then
$A\circ B=KAB$ and $B\circ A=BKA$, so that
\be
[A,B]_\circ=[KA,B].
\ee
Thus the dynamics described by the Heisenberg operator $H$ as above
is the same as the dynamics described by $K^{-1}H$ with respect to the 
new product. This time, however, the new product is not compatible
with the standard one and we have $a^\da\circ a=0$.

\section{Conclusions}
We have shown that in  the  Heisenberg  picture  alternative  associative
products are possible which allow to describe the   same   dynamics   on
the  space  of observables. We hope to consider the corresponding version
on the phase space via the Wigner map, to compare these findings with
those available at the classical level.

\end{document}